\documentclass[useAMS,usenatbib]{mn2e}
\usepackage{graphicx}
\title[Lithium Deficit in Stars with Planets]{Parent Stars of Extrasolar Planets - XIV. Strong Evidence of Li Abundance Deficit}
\author[G.\ Gonzalez]{G.\ Gonzalez$^{1}$\\
$^{1}$Department of Physics and Astronomy, Ball State University, Muncie, IN 47306 USA\\
}

\begin{document}

\date{Accepted ??. Received ??; in original form ??}

\pagerange{\pageref{firstpage}--\pageref{lastpage}} \pubyear{??}

\maketitle

\label{firstpage}

\begin{abstract}
We report the results of our analysis of new high resolution spectra of 30 late-F to early-G dwarf field stars for the purpose of deriving their Li abundances. They were selected from the subsample of stars in the Valenti and Fischer  compilation that are lacking detected planets. These new data serve to expand our comparison sample used to test whether stars with Doppler-detected giant planets display Li abundance anomalies. Our results continue to show that Li is deficient among stars with planets when compared to very similar stars that lack such planets. This conclusion is strengthened when we add literature data to ours in a consistent way. We present a table of stars with planets paired with very similar stars lacking planets, extending the recent similar results of Delgado Mena et al.
\end{abstract}

\section{Introduction}

In this study we again revisit the question of a possible correlation between the presence of Doppler-detected planets and host star Li abundance. Several studies \citep{is04,tk05,gg08,is09,sou10,gg10,del14,fig14} indicate that stars with planets (SWPs) have lower Li abundances compared to stars without detected planets near the solar temperature. However, other studies \citep{ryan00,lh06,bau10,gh10,ram12} have failed to confirm this pattern. Therefore, despite having received attention for over a decade from several inpedendent research groups, this question remains controversial.

The present work is very similar to \citet{gg14}, but we increases the size of the sample of comparison stars with new spectroscopic data and also add additional literature data. In \citet{gg14} we showed that SWPs in the temperature (T$_{\rm eff}$) range $5600 <$ T$_{\rm eff} < 5800$ K are deficient in Li by about 0.5 dex relative to comparison stars with similar properties. That study also revealed weaker evidence that SWPs with T$_{\rm eff} > 6100$ K are deficient in Li. However, relatively few comparison stars with T$_{\rm eff} < 5600$ K and T$_{\rm eff} > 6100$ K were employed in that study. We have targeted comparison stars in these ranges in the present work.

The purpose of the present study is to test again the claim that the Li abundances of Sun-like SWPs are different than those of similar stars without known planets. In Section 2 we describe our new spectroscopic observations and Li abudance analyses. In Section 3 we compare SWPs and stars without detected planets. We discuss the results in Section 4 and present the conclusions in Section 5.

\section{Observations and analyses}

We observed 30 stars without detected planets selected from the \citet{vf05} study of field dwarfs. Only one of them, HD 162826, was already included as a comparison star in one of our previous studies \citep{gg10}; it was observed on two nights. None of the other stars in our program have previously determined Li abundances. The values of T$_{\rm eff}$ listed by \citet{vf05} for our program stars range from 5630 to 6274 K, and the average is near 5950 K. We observed our target stars on May 9-12, 2014 using the McDonald Observatory 2.1-m Otto Struve telescope and Sandiford spectrograph, which is a Cassegrain echelle design \citep{mcc93}. The instrument setup, observing procedures and analysis steps are nearly identical to those used in our previous observing run for this program and are described in \citet{gg14}. A solar spectrum was obtained via reflected light off the Galilean moon Ganymede, and the hot star Regulus was observed in order to correct for telluric features.

As in \citet{gg14} we reduced the spectra in IRAF and measured the equivalent widths in an automated way using DAOSPEC \citep{sp08}. The stellar atmosphereic parameters and Li abundances were calculated using MOOG. We used the same linelist and atomic parameters as in \citet{gg14}. We list the results of our Fe line analysis in Table 1.

Unfortunately, we were only able to set upper limits on the Li abundance for most of the cooler stars in our new sample. Most of the Li detections are for stars hotter than T$_{\rm eff} \simeq 5700$ K.

Comparing our results to those of \citet{vf05}, we find that the mean $\Delta$T$_{\rm eff} = 20 \pm 67$ K and the mean $\Delta$log g $= 0.02 \pm 0.15$ dex (in the sense of our values minus theirs). Both offsets are small, and in each case the scatter of the differences is consistent with the uncertainties quoted in the studies. As noted above, there is one star in common between the present work and our previous studies, HD 162826. The stellar parameters determined in the two studies are consistent with each other within the quoted uncertainties. For the purpose of the analysis presented below in Section 3, we will use a weighted average set of parameters determined for this star.

The derived parameters for each star are listed in the last three columns of Table 1. They were determined using the procedure described in \citet{gg14}. The mean difference between our spectroscopic log g values and the parallax-derived (photometric) values is 0.06 $\pm$ 0.10 dex. Within the quoted error, the photometric log g values are consistent with the spectroscopic values.

Only one star from the present work, HD 162826, is included in \citep{ram12}, which is a compilation of Li abundances for 1381 FGK dwarf and subgiant stars determined by them and also drawn from the literature.

\begin{table*}
\centering
\begin{minipage}{160mm}
\caption{Parameters of the program stars determined from our spectroscopic analyses. Derived parameters based on stellar isochrones are given in columns 8 to 10.}
\label{xmm}
\begin{tabular}{rrcccccccc}
\hline
Star & & T$_{\rm eff}$ & log g & $\zeta_{\rm t}$ & [Fe/H] & log $\epsilon$(Li) & mass (M$_{\odot}$) & log g & age (Gyr)\\
HD & HIP & (K) & & (km~s$^{\rm -1}$) & & & & & \\
\hline
73344 & 42403 & $6172\pm52$ & $4.48\pm0.05$ & $1.59\pm0.12$ & $~~0.14\pm0.04$ & $~~~2.55\pm0.08$ & $1.20\pm0.02$ & $4.37\pm0.02$ & $0.8\pm0.6$\\
73668 & 42488 & $5974\pm39$ & $4.47\pm0.06$ & $1.51\pm0.12$ & $~~0.01\pm0.03$ & $~~~2.40\pm0.07$ & $1.08\pm0.02$ & $4.38\pm0.04$ & $2.8\pm1.7$\\
76752 & 44089 & $5749\pm32$ & $4.32\pm0.03$ & $1.32\pm0.08$ & $~~0.02\pm0.02$ & $<1.21\pm0.10$ & $0.99\pm0.02$ & $4.30\pm0.03$ & $8.1\pm1.1$\\
85301 & 48423 & $5728\pm47$ & $4.56\pm0.05$ & $1.40\pm0.08$ & $~~0.13\pm0.04$ & $~~~2.02\pm0.08$ & $1.02\pm0.02$ & $4.49\pm0.02$ & $0.8\pm0.7$\\
87836 & 49680 & $5740\pm47$ & $4.28\pm0.07$ & $1.31\pm0.07$ & $~~0.30\pm0.03$ & $~~~1.54\pm0.08$ & $1.11\pm0.03$ & $4.25\pm0.03$ & $5.5\pm1.0$\\
88371 & 49942 & $5676\pm34$ & $4.33\pm0.06$ & $1.35\pm0.13$ & $-0.27\pm0.02$ & $<0.88\pm0.10$ & $0.87\pm0.01$ & $4.33\pm0.03$ & $11.3\pm0.4$\\
88986 & 50316 & $5798\pm44$ & $4.09\pm0.06$ & $1.58\pm0.08$ & $~~0.00\pm0.03$ & $~~~1.80\pm0.07$ & $1.06\pm0.02$ & $4.08\pm0.02$ & $8.0\pm0.5$\\
91204 & 51579 & $5978\pm40$ & $4.33\pm0.07$ & $1.53\pm0.08$ & $~~0.24\pm0.03$ & $~~~2.06\pm0.07$ & $1.17\pm0.02$ & $4.35\pm0.03$ & $1.9\pm1.1$\\
96418 & 54347 & $6290\pm59$ & $4.26\pm0.11$ & $2.52\pm0.25$ & $-0.04\pm0.04$ & $<1.72\pm0.10$ & $1.30\pm0.05$ & $4.05\pm0.03$ & $3.2\pm0.5$\\
98388 & 55262 & $6297\pm66$ & $4.32\pm0.07$ & $2.28\pm0.22$ & $~~0.06\pm0.05$ & $~~~2.73\pm0.09$ & $1.22\pm0.03$ & $4.32\pm0.03$ & $1.3\pm0.9$\\
101472 & 56960 & $6252\pm52$ & $4.45\pm0.07$ & $1.81\pm0.20$ & $-0.05\pm0.04$ & $~~~3.07\pm0.08$ & $1.12\pm0.02$ & $4.43\pm0.02$ & $0.5\pm0.4$\\
103432 & 58067 & $5678\pm38$ & $4.55\pm0.07$ & $1.14\pm0.10$ & $-0.09\pm0.03$ & $<1.23\pm0.10$ & $0.94\pm0.03$ & $4.49\pm0.04$ & $2.8\pm2.5$\\
107213 & 60098 & $6333\pm57$ & $4.22\pm0.07$ & $2.22\pm0.12$ & $~~0.24\pm0.04$ & $<1.61\pm0.10$ & $1.54\pm0.02$ & $3.96\pm0.03$ & $2.1\pm0.1$\\
107705 & 60353 & $6202\pm52$ & $4.40\pm0.05$ & $1.60\pm0.14$ & $~~0.17\pm0.04$ & $~~~2.76\pm0.08$ & $1.22\pm0.02$ & $4.35\pm0.02$ & $0.8\pm0.6$\\
110537 & 62039 & $5711\pm39$ & $4.41\pm0.08$ & $1.31\pm0.08$ & $~~0.09\pm0.03$ & $<1.32\pm0.12$ & $1.00\pm0.02$ & $4.35\pm0.04$ & $6.4\pm1.9$\\
126961 & 70782 & $6148\pm59$ & $4.30\pm0.07$ & $1.67\pm0.17$ & $~~0.09\pm0.04$ & $~~~2.80\pm0.08$ & $1.18\pm0.03$ & $4.32\pm0.03$ & $2.0\pm1.2$\\
127334 & 70873 & $5635\pm50$ & $4.11\pm0.06$ & $1.22\pm0.17$ & $~~0.19\pm0.04$ & $<1.29\pm0.12$ & $1.02\pm0.02$ & $4.25\pm0.03$ & $8.6\pm1.3$\\
129814 & 72043 & $5785\pm37$ & $4.20\pm0.05$ & $1.41\pm0.11$ & $-0.06\pm0.03$ & $<1.10\pm0.10$ & $0.97\pm0.02$ & $4.27\pm0.03$ & $9.0\pm1.0$\\
136580 & 75039 & $6286\pm59$ & $4.50\pm0.08$ & $2.19\pm0.26$ & $-0.10\pm0.04$ & $~~~2.70\pm0.08$ & $1.15\pm0.03$ & $4.26\pm0.03$ & $3.2\pm0.8$\\
138776 & 76228 & $5629\pm37$ & $4.16\pm0.08$ & $1.26\pm0.05$ & $~~0.32\pm0.03$ & $<0.94\pm0.10$ & $1.05\pm0.02$ & $4.30\pm0.07$ & $6.0\pm2.2$\\
142267 & 77801 & $5823\pm52$ & $4.47\pm0.07$ & $1.39\pm0.19$ & $-0.38\pm0.04$ & $<1.12\pm0.10$ & $0.86\pm0.02$ & $4.40\pm0.03$ & $8.9\pm2.0$\\
147231 & 79619 & $5632\pm47$ & $4.44\pm0.07$ & $1.13\pm0.11$ & $~~0.01\pm0.04$ & $<0.84\pm0.10$ & $0.94\pm0.02$ & $4.36\pm0.03$ & $8.6\pm2.0$\\
149200 & 81062 & $6292\pm52$ & $4.35\pm0.07$ & $2.01\pm0.14$ & $~~0.11\pm0.04$ & $~~~2.27\pm0.08$ & $1.28\pm0.02$ & $4.21\pm0.03$ & $2.3\pm0.4$\\
152446 & 82568 & $6242\pm66$ & $4.27\pm0.08$ & $2.00\pm0.23$ & $-0.09\pm0.05$ & $~~~1.96\pm0.09$ & $1.25\pm0.05$ & $4.04\pm0.03$ & $3.6\pm0.6$\\
154160 & 83435 & $5488\pm42$ & $3.87\pm0.05$ & $1.32\pm0.06$ & $~~0.29\pm0.03$ & $~~~1.25\pm0.08$ & $1.16\pm0.02$ & $3.96\pm0.02$ & $6.7\pm0.4$\\
155423 & 84082 & $6275\pm66$ & $4.41\pm0.08$ & $1.96\pm0.14$ & $~~0.29\pm0.05$ & $~~~2.20\pm0.09$ & $1.33\pm0.02$ & $4.24\pm0.03$ & $1.5\pm0.6$\\
162826 & 87382 & $6186\pm52$ & $4.35\pm0.07$ & $1.80\pm0.17$ & $~~0.02\pm0.04$ & $~~~2.47\pm0.07$ & $1.17\pm0.02$ & $4.27\pm0.03$ & $3.0\pm0.8$\\
162826 & 87382 & $6165\pm59$ & $4.36\pm0.10$ & $1.64\pm0.18$ & $~~0.03\pm0.04$ & $~~~2.43\pm0.08$ & $1.17\pm0.03$ & $4.27\pm0.03$ & $3.2\pm0.9$\\
176377 & 93185 & $5910\pm45$ & $4.56\pm0.07$ & $1.40\pm0.16$ & $-0.19\pm0.04$ & $~~~2.08\pm0.07$ & $0.98\pm0.03$ & $4.48\pm0.02$ & $1.7\pm1.4$\\
181144 & 94905 & $6323\pm59$ & $4.57\pm0.05$ & $1.77\pm0.22$ & $-0.03\pm0.04$ & $~~~3.03\pm0.08$ & $1.17\pm0.02$ & $4.38\pm0.02$ & $0.7\pm0.6$\\
189067 & 98206 & $5829\pm46$ & $4.10\pm0.07$ & $1.45\pm0.12$ & $-0.05\pm0.04$ & $~~~1.83\pm0.07$ & $1.03\pm0.02$ & $4.10\pm0.03$ & $8.3\pm0.6$\\
\hline
\end{tabular}
\end{minipage}
\end{table*}

\section{Comparison of samples}
\subsection{New data}

We formed our McDonald SWP and comparison stars samples by combining the new results in the present work and the McDonald data results from \citet{gg14} and ealier papers in this series. In this section we limit our comparison of Li abundances between SWPs and stars without planets to T$_{\rm eff} =$ 5650 to 6350 K. Also, stars having only upper limits for the Li abundances are not included in the analysis. These selection criteria leave us with 55 SWPs (no additional SWPs have been added in the present work) and 85 comparison stars (68 from previous work and an additional 17 from the present work).

We calculated a weighted-average Li abundance difference between each SWP and all the comparison stars using $(\Delta_{\rm p,c})^{-2}$ as the weight, where $\Delta_{\rm p,c}$ is is a measure of the distance between two stars in T$_{\rm eff}$-[Fe/H]-$\log g$-$\log$ age space; the calculation of this quantity is described in detail in \citet{gg14}.\footnote{Note, in preparing Figures 8 to 13 in \citet{gg14} we had inadvertently calculated $\Delta_{\rm p,c}$ using `age' instead of `$\log$ age'. This error has been corrected in the present work.} We show the weighted average Li abundance differences between the SWPs and comparison stars in Figure 1.

To correct for bias present in Figure 1 we follow the same procedure employed with Figures 6, 7, 11 and 12 of \citet{gg14}. First, we used the 85 comparison stars to produce Figure 2 and calculated the average Li abundance differences in bins 100 K wide. These average difference corrections were then applied to the data in Figure 1 to produce the corrected data, shown in Figure 3.

\begin{figure}
  \includegraphics[width=3in]{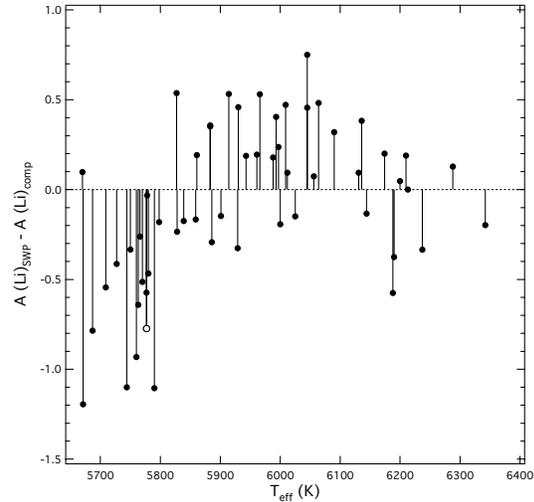}
 \caption{Weighted average Li abundance differences between SWPs and comparison stars. The open circle represents the Sun.}
\end{figure}

\begin{figure}
  \includegraphics[width=3in]{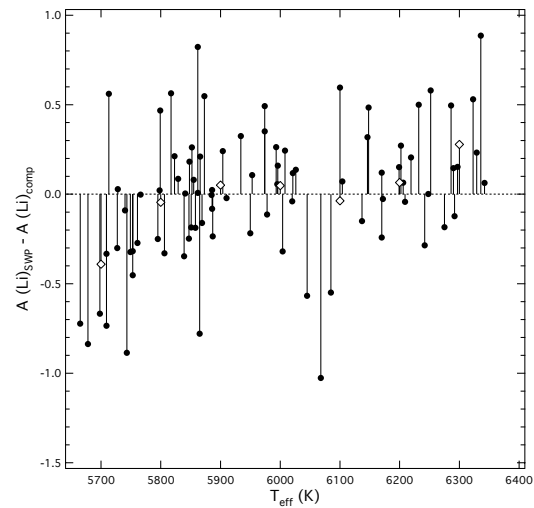}
 \caption{Weighted average Li abundance differences among the comparison stars. Average differences in intervals of 100 K are shown as diamond symbols.}
\end{figure}

\begin{figure}
  \includegraphics[width=3in]{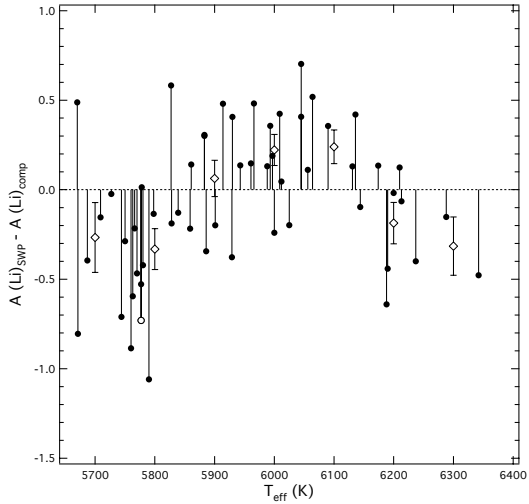}
 \caption{Same data as shown in Figure 1 but corrected for bias using the average corrections from Figure 2.}
\end{figure}

\subsection{Literature dataset}

In \citet{gg14} we added data from the peer-reviewed literature in order to test the robustness of the results obtained using only the McDonald data. We follow a similar procedure here. In \citet{gg14} we combined the new data acquired in that work to the large literature compilation of \citet{ram12}. \citet{gg14} calculated offsets to their data by comparing stars in common with \citet{ram12}, thereby placing the McDonald-based results on the same scale as \citet{ram12}. We add the new data in the present work to the ``literature dataset'' of \citet{gg14} following the same procedure.

The literature dataset of \citet{gg14} consists of 100 SWPs and 241 non-SWPs. We applied the same corrections to our new data as we did in \citet{gg14} prior to adding them to the new literature dataset.\footnote{Two stars from the literature dataset in \citet{gg14} was mistakenly listed by \citet{ram12} as having a planet, HIP 35265 and 55868. We have moved these stars from the SWP to the non-SWP category in the present work.}

We also added data from \citet{del14} to the new literature dataset. These data were not included in \citet{gg14}. \citet{del14} is a spectroscopic analysis of 326 Sun-like stars with and without planets having $5600 <$ T$_{\rm eff} < 5900$ K. We selected stars from the online version of their Table 5, which lists the stars in their sample without detected planets; we retained 139 stars with detected Li for further analysis. We also added 19 SWPs from their Tables 4 and 6 with Li detections and planets with minimum masses greater than Neptune.

Among the stars we selected from \citet{del14}, 41 are also present in our literature dataset. The differences (in the sense \citet{del14} minus \citet{ram12}) in T$_{\rm eff}$, $\log g$, [Fe/H], and Li abundance are $32 \pm 17$ K, $0.06 \pm 0.06$ dex, $0.02 \pm 0.02$ dex, and $0.00 \pm 0.07$ dex, respectively (where the errors are simple standard deviations). We have applied these differences to the \citet{del14} data prior to adding them to the literature dataset. In addition, the average age of the \citet{del14} stars is $0.8 \pm 0.3$ of that of the literature dataset stars; we corrected the ages of the \citet{del14} stars by dividing their ages by this ratio prior to adding them to the literature dataset. For those cases where there is more than one measurement, we calculated a weighted average, where the weight is based on the number of measurements of the star. Our final version of the new literature dataset consists of 108 SWPs and 358 non-SWPs.

We show the weighted average Li abundance differences from the literature dataset in Figure 4. The overal pattern is similar to that in Figure 1. Figure 5 shows the Li abundance differences among the comparison stars in the literature dataset. As we did with the data in plotted in Figure 2, we calculated the average bias correction for each 100 K wide bin in T$_{\rm eff}$. These average values were used to correct the data in Figure 4. The resulting bias-corrected weighted Li abundance differences are shown in Figure 6. Also shown in Figure 6 are the average difference in each T$_{\rm eff}$ bin along with the standard deviation of the average.

\begin{figure}
  \includegraphics[width=3in]{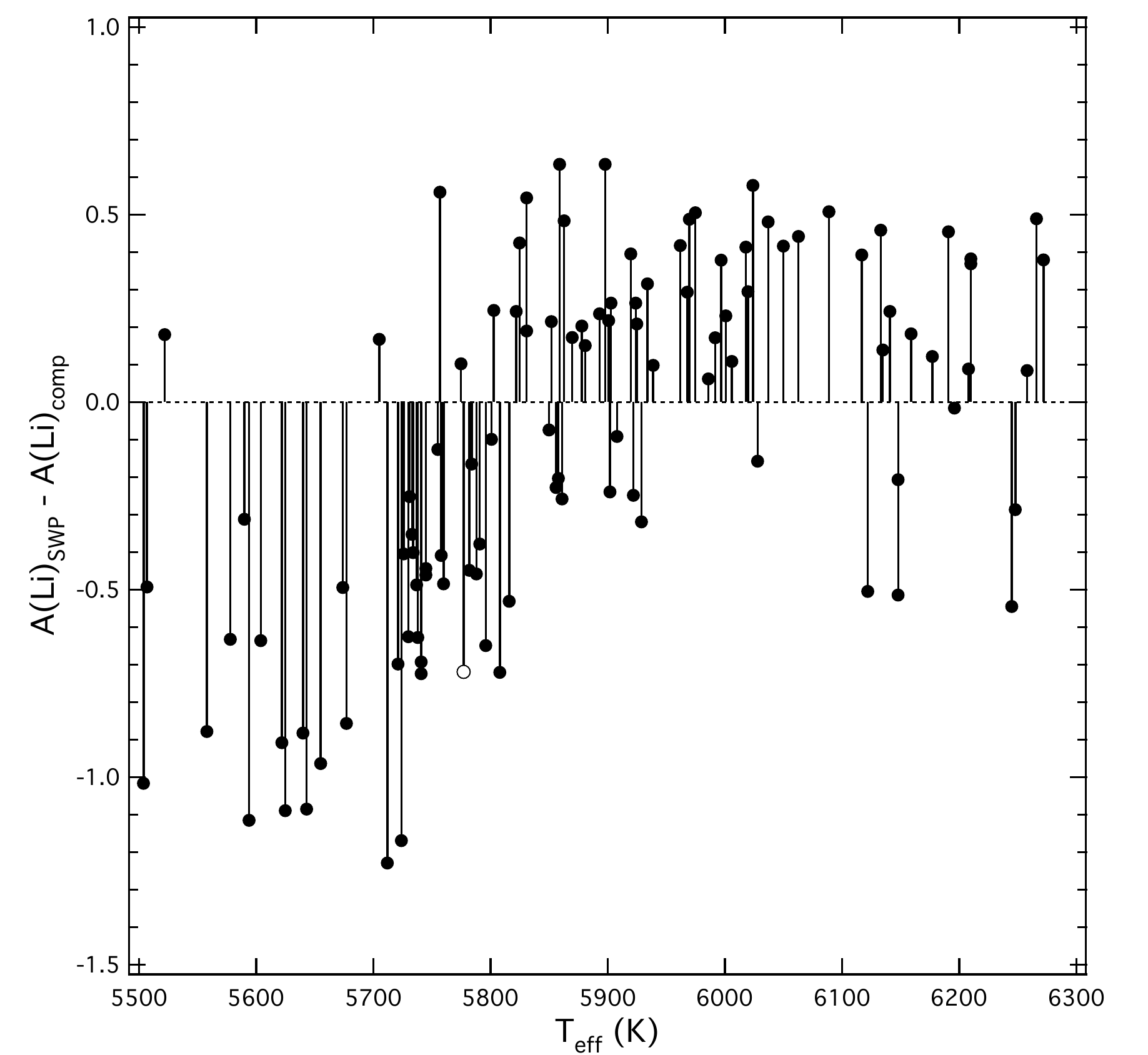}
 \caption{Weighted average Li abundance differences between the SWPs and comparison stars from the literature dataset.}
\end{figure}

\begin{figure}
  \includegraphics[width=3in]{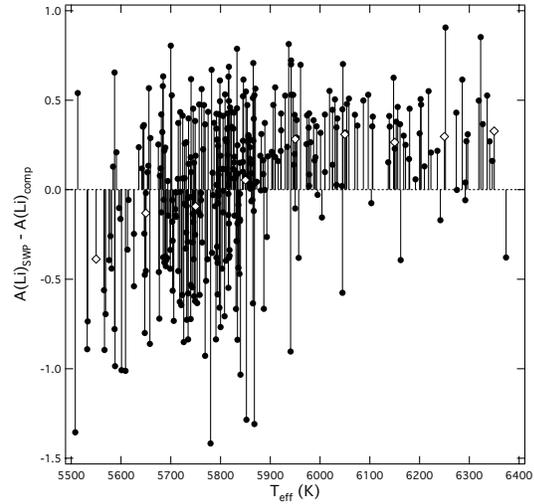}
 \caption{Weighted average Li abundance differences among the comparison stars from the literature dataset. The calculations were done in the same way as in Figure 4. The open diamonds are the averages of the Li abundance differences in 100 K wide bins.}
\end{figure}

\begin{figure}
  \includegraphics[width=3in]{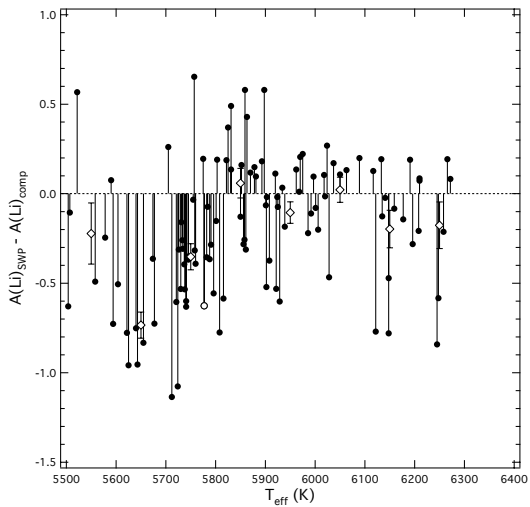}
 \caption{Data from Figure 4 corrected for bias using the binned bias corrections from Figure 5. The open diamonds represent the averages of the Li abundance differences in 100 K-wide bins, and the error bars correspond to the standard deviation of the average.}
\end{figure}

The simplest way to compare the Li abundances in the SWP and non-SWP samples is to find the smallest difference in the $\Delta_{p,c}$ index between each SWP and a non-SWP comparison star; the Li abundance difference is then calculated between the SWP and the non-SWP star with the smallest $\Delta_{p,c}$ value. We first applied this analysis method in \citet{gg14}\footnote{Note, in repeating these calculations with the new data, we discovered an error in the computer code used in this analysis, which prevented each SWP from being compared to the full sample of comparison stars. This error affected Figure 13 of \citet{gg14}. In addition, the same error noted earlier about using `age' instead of `$\log$ age' in the calculation of the $\Delta_{p,c}$ values also affected this figure. Both errors have been corrected in the present work.}, and we repeat it here with the same dataset used to prepare Figures 4 to 6. The results are shown in Figure 7. No bias corrections were applied, as it is unlikely that they should be needed in this case. The overall pattern of the average Li abundances differences in Figure 7 is similar to that of Figure 6.

The average Li abundance difference for all the SWPs plotted in Figures 6 and 7 are $-0.19\pm0.04$ and $-0.17\pm0.04$ (s.e.m.), respectively. Thus, the mean of the Li abundance among the SWPs differs from the non-SWPs by about 4$\sigma$ to nearly 5$\sigma$. It appears by visual inspection of Figures 6 and 7 that the number of SWPs with negative Li abundance differences is greater than those with positive differences. We can verify this by counting stars with positive and negative deviations. In Figure 6 the number of positive and negative Li differences are 38 and 70, respectively. In Figure 7, the number of positive and negative Li differences are 36 and 72, respectively. In each of Figures 6 and 7 all but two of the eight binned mean differences are significantly less than zero.

\begin{figure}
  \includegraphics[width=3in]{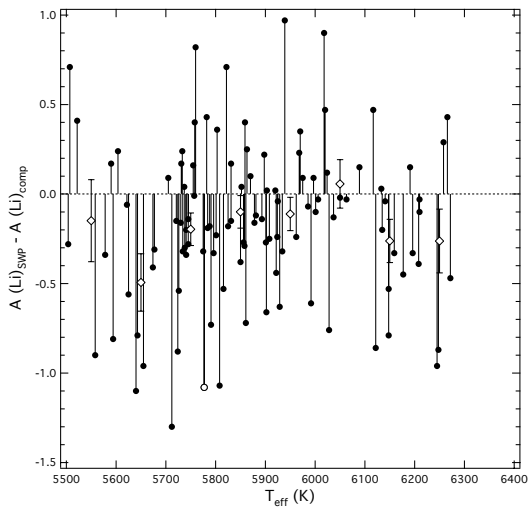}
 \caption{Li abundance difference between each SWP and the most similar comparison star. Symbols have same meanings as in Figure 6.}
\end{figure}

\section{Discussion}

\citet{pace12} showed that there exists a large scatter in Li abundance among stars of a given mass for mass $\le$ 1.1 M$_{\odot}$ in the open cluster M67. Now that giant planets have been detected in M67 \citep{bru14}, the Li abundances can be compared between the SWPs and non-SWPs in the cluster. Two of the stars with giant planets in M67 are G dwarfs. One, YBP 1194, is a probable solar analog and has a Li abundance between -0.2 dex \citep{pas08} and 0.2 dex  \citep{one11} relative to solar. From examination of Figure 2 of \citet{pace12} we can see that several cluster members with similar mass have larger Li abundances than YBP 1194. Certainly, additional high resolution spectroscopy of the stars being searched for planets in M67 are warranted for the purpose of determining accurate Li abundances.

\citet{del14} give multiple examples of nearby stars that are indistinguishable in their measured parameters (within the measurement errors), yet have very different Li abundances; they compare 15 very similar pairs of stars in their Figure 10, which shows this phenomenon.\footnote{\citet{del14} differs from our present work in that they also include stars with only upper limits on their Li abundance.} This implies that a parameter in addition to the ones already known to influence Li abundance (T$_{\rm eff}$, mass, metallicity, age) must be involved. To explore this further, we list in Table 2 (complete version online) the 108 SWPs from our literature dataset and the most similar non-SWP for each SWP. We confirm that indeed there are many examples of pairs of nearly identical stars with significantly different Li abundances; in comparing the pairs of stars in Table 2 it is important to note that the typical uncertainty in the Li abundance is about 0.10 dex. For many of the pairs the difference in Li abundance is much greater than the measurement uncertainty.

Many of the pairs listed in Table 2 have similar Li abundances, but there are several cases wherein a SWP has an order-of-magnitude smaller Li abundance than its non-SWP twin. Examples of these SWPs include the Sun, HIP 17960, HIP 1499, HIP 20723, HIP 31540, HIP 32916, HIP 36795, HIP 39417, HIP 52409, and HIP 89844. In addition, there are a few cases wherein the Li abundance of a SWP is much larger than its twin; these include HIP 40687, HIP 42723, and HIP 95262. Each of these pairs deserves more detailed study.

The average value of m$_{\rm p}$sini for the SWPs in the literature dataset is 2.5 M$_{\rm J}$, and it ranges from 0.024 to 17.4 M$_{\rm J}$. \citet{del14} present weak evidence that Li depletion is greater among SWPs with more massive planets. We can test this with our literature dataset. The average Li abundance difference between an SWP and its non-SWP twin when m$_{\rm p}$sini $> 1$ M$_{\rm J}$ is $-0.14\pm 0.05$; for less massive planets it is $-0.22\pm 0.07$.  This difference is not significant.

Although the reality of a deficit of Li among SWPs compared to non-SWPs appears to be now well-established, there is still room for improvements. The analysis can still benefit from more comparison non-SWPs. In addition, any improvements in the determination of the fundamantal parameters of the SWPs and non-SWPs will lead to more accurate $\Delta_{p,c}$ values and, therefore, weights in the analysis.

\begin{table*}
\centering
\begin{minipage}{160mm}
\caption{The SWPs from the literature dataset as well as the most similar non-SWP for each SWP. The complete table is available as on online supplement; the online data also lists the estimated age of each star.}
\label{xmm}
\begin{tabular}{lcccclccccc}
\hline
SWP & & & & & non-SWP & & & & &\\
HIP & T$_{\rm eff}$ & log g & [Fe/H] & log Li & HIP & T$_{\rm eff}$ & log g & [Fe/H] & log Li & $\Delta_{p,c}$\\
\hline
0 (Sun) & 5777 & 4.44 & 0.00 & 1.10 & 97420 & 5792 & 4.43 & 0.02 & 2.18 & 0.06\\
80 & 5862 & 4.30 & -0.55 & 1.80 & 13267 & 5845 & 4.26 & -0.59 & 1.77 & 0.11\\
\hline
\end{tabular}
\end{minipage}
\end{table*}

\section{Conclusions}

We present the results of our analysis of high quality spectra of 30 F and G dwarfs, including Li abundances. The new stars analyzed in this work were selected, for the most part, to increase the number of comparison stars with T$_{\rm eff} \sim 6200$ K. When combined with a large homogeneous sample of similar stars with Li abundance determinations from the literature, we were able to confirm that the Li abundances of SWPs with T$_{\rm eff} \sim 5700$ K are significantly smaller than those of stars without detected planets; there is weaker evidence that Li is also deficient among SWPs with T$_{\rm eff} \sim 6200$ K. Lithium is deficient among all the SWPs in our literature dataset by 4$\sigma$ to 5$\sigma$, depending on how the samples are compared. Our results generally confirm other recent independent studies of Li abundances in SWPs \citep{tk10,del14,fig14}.

Like \citet{del14}, we have drawn attention to SWP-non-SWP pairs with very similar properties, yet with very different Li abundances. These cases are very unlikely to be explained by observational error. They deserve additional detailed study.

\section*{Acknowledgments}

We thank the anonymous reviewer for helpful comments.

\bsp

\label{lastpage}

\end{document}